\documentclass{PoS}

\title{{\footnotesize\vspace*{-3.cm}\hspace*{9.5cm} CERN-PH-TH/2012-309, MZ-TH/12-48}\vspace*{3.cm}\\Flavour data constraints on new physics and SuperIso}

\ShortTitle{Flavour data constraints on new physics and SuperIso}

\author{\speaker{Farvah Mahmoudi}\\
	CERN Theory Division, CH-1211 Geneva 23, Switzerland\\
        Clermont Universit\'e, Universit\'e Blaise Pascal, CNRS/IN2P3, LPC, BP 10448, F-63000 Clermont-Ferrand, France\\
        E-mail: \email{mahmoudi@in2p3.fr}}

\author{Tobias Hurth\\
	PRISMA Cluster of  Excellence \& Institute for Physics (THEP),
	Johannes Gutenberg University, D-55099 Mainz, Germany\\
	CERN Theory Division, CH-1211 Geneva 23, Switzerland\\
E-mail: \email{tobias.hurth@cern.ch}}

\abstract{
We discuss the implications of $B_s\to\mu^+\mu^-$ and $B\to K^*\mu^+\mu^-$ decays in the context of indirect searches for new physics, emphasising the new LHCb results.
In particular, we derive the consequences of the MFV hypothesis and discuss the importance of the MFV predictions. The impact of the recent LHCb measurements in the context of the MSSM will also be addressed, and the SuperIso program will be briefly described.  
}

\FullConference{36th International Conference on High Energy Physics\\
         4-11 July 2012\\
         Melbourne, Australia}

\begin{document}

\section{Introduction}

The importance and deep implications of the information from flavour physics in the search for new physics (NP) is very well known. The flavour sector of the Standard Model (SM) provides tests of the quantum structure of the SM at loop level, of the Minimal Flavour Violation (MFV) hypothesis, and can probe sectors inaccessible to direct searches. In the context of specific NP scenarios such as supersymmetry (SUSY), indirect information from flavour physics is complementary to the direct search information and very strong constraints on the SUSY parameter space can be obtained.

While new physics particles and signatures are searched for very actively in ATLAS and CMS experiments, the LHCb experiment has also a rich BSM program through indirect searches. 
The key processes here are $B_s \to \mu^+ \mu^-$ and $B \to K^* \mu^+ \mu^-$, in addition to the CP violating processes (not covered here). Given the impressive progress in the experimental accuracies, it is now crucial to have a precise and clear estimation of the SM predictions and errors in order to deduce solid constraints on the NP parameters.

\section{Theoretical framework}
The effective Hamiltonian is the starting point for the calculation of the flavour observables: 
\begin{equation}
{\cal H}_{\rm eff}  =  -\frac{4G_{F}}{\sqrt{2}} V_{tb} V_{ts}^{*} \, \Bigl(\,\sum_{i=1\cdots10} \bigl(C_{i}(\mu) O_i(\mu)+C'_{i}(\mu) O'_i(\mu)\bigr)\Bigr)\;,
\end{equation}
where $O_i(\mu)$ denote the operators and $C_i(\mu)$ their corresponding Wilson coefficients evaluated at the scale $\mu$ encoding the short distance physics.
The primed operators are chirality flipped compared to the non-primed operators and are highly suppressed in the SM.
In general, contributions from physics beyond the SM to the observables can be described by the modification of Wilson coefficients and/or by the addition of new operators.

\section{The decays $B_{s} \to \mu^+ \mu^-$ and $B \to K^* \mu^+ \mu^-$}
The rare leptonic and semileptonic decays $B_s\to\mu^+\mu^-$ and $B\to K^*\mu^+\mu^-$ deserve special attention as they are high priority observables for the LHCb experiment, and updated results were obtained recently. The decay $B_{s} \to \mu^+ \mu^-$ is not yet observed but a stringent 95\% C.L. limit on the branching ratio BR$(B_s\to\mu^+\mu^-) < 4.5\times 10^{-9}$ has been reported \cite{Aaij:2012ac}. 
The theoretical prediction of this branching ratio can be obtained using \cite{Bobeth:2001sq,superiso}:
\begin{eqnarray}
  \label{eq:Bs2mm_formula}
\mbox{BR}(B_s\to\mu^+\mu^-)&=&\frac{G_F^2 \alpha^2}{64\pi^2}f_{B_s}^2
m_{B_s}^3 |V_{tb}V_{ts}^*|^2\tau_{B_s}\sqrt{1-\frac{4m_\mu^2}{m_{B_s}^2}}\\
&&\times\left\{\left(1-\frac{4m_\mu^2}{m_{B_s}^2}\right)
  |C_{Q_1}-C'_{Q_1}|^2+\left|(C_{Q_2}-C'_{Q_2})+2(C_{10}-C'_{10})\frac{m_\mu}{m_{B_s}}\right|^2\right\}\,.\nonumber  
\end{eqnarray}
In the SM, the only non vanishing Wilson coefficient is $C_{10}$ which receives contributions from $Z$ penguin and box diagrams. Using the input parameters given in \cite{Mahmoudi:2012un} we obtain BR$(B_s\to\mu^+\mu^-)_{SM}=(3.53\pm 0.38)\times 10^{-9}$. The global uncertainty is about 11\% with the largest error from the lattice evaluation of $f_{B_s}$. As the experimental limit is very close to the SM prediction, large contributions from NP are not allowed anymore.

The decay $B\to K^*\mu^+\mu^-$ has the advantage of offering a variety of complementary observables.
The differential decay distribution of the $\bar B ^0 \to \bar K ^*(\to K^- \pi^+ ) \mu^+ \mu^-$ decay is expressed in terms of three angles $\theta_l$, $\theta_{K^*}$, $\phi$ and 
the invariant dilepton mass squared ($q^2$) \cite{Kruger:2005ep}:
\begin{equation}\label{eq:diffAD}
  d^4\Gamma = \frac{9}{32\pi} J(q^2, \theta_l, \theta_{K^*}, \phi)\, dq^2\, d\cos\theta_l\, d\cos\theta_{K^*}\, d\phi \;,
\end{equation}
where $J(q^2, \theta_l, \theta_{K^*}, \phi)$ can be expanded in terms of the angular coefficients $J_i$ described in turn in terms of the transversity amplitudes and form factors \cite{Beneke:2001at}.
The dilepton mass distribution can be obtained by integrating Eq.~(\ref{eq:diffAD}) over all angles \cite{Bobeth:2008ij}:
\begin{equation}
\frac{d\Gamma}{dq^2} = \frac{3}{4} \bigg( J_1 - \frac{J_2}{3} \bigg)\;.
\label{eq:dBR}
\end{equation}
The forward-backward asymmetry $A_{FB}$ benefits from reduced theoretical uncertainty and can be defined as:
\begin{equation}
A_{\rm FB}(q^2)  \equiv
     \left[\int_{-1}^0 - \int_{0}^1 \right] d\cos\theta_l\, 
          \frac{d^2\Gamma}{dq^2 \, d\cos\theta_l} \Bigg/\frac{d\Gamma}{dq^2}
          =  -\frac{3}{8} J_6 \Bigg/ \frac{d\Gamma}{dq^2}\; .
\label{eq:AFB}
\end{equation}
The zero--crossing of the forward-backward asymmetry ($q_0^2$) is of special interest as the form factors cancel out at leading order. Moreover, $q_0^2$ is sensitive to the relative sign of $C_7$ and $C_9$ and therefore its measurement allows to remove the sign ambiguity. 

From the ratio of the transversity amplitudes one can also construct the longitudinal polarisation fraction $F_L$, which reads:
\begin{equation}
 F_L(s) = \frac{-J_2^c}{d\Gamma / dq^2}\;.
\end{equation}
In Table~\ref{tab:experiment} we summarise the SM predictions and experimental values for these observables.
\begin{table}[!t]
\begin{center}
\begin{tabular}{|l|l|l|l|l|l|l|}\hline 
  Observable                                                                & SM prediction & Experiment       \\ \hline \hline
  $10^7 \mbox{GeV}^2 \times \langle dBR/dq^2\; (B \to K^* \mu^+ \mu^-) \rangle_{[1,6]}$ & $0.47 \pm 0.27 $        & $0.42 \pm 0.04 \pm 0.04$   \\ \hline
  $\langle A_{FB}(B \to K^* \mu^+ \mu^-) \rangle_{[1,6]}$         & $-0.06 \pm 0.05 $        & $-0.18 ^{+0.06+0.01}_{-0.06-0.01}$   \\ \hline
  $\langle F_{L}(B \to K^* \mu^+ \mu^-) \rangle_{[1,6]}$          & $0.71 \pm 0.13 $        & $0.66 ^{+0.06+0.04}_{-0.06-0.03}$   \\ \hline
  $q_0^2 (B \to K^* \mu^+ \mu^-)/\mbox{GeV}^2$      & $4.26 ^{+0.36}_{-0.34} $        & $4.9  ^{+1.1}_{-1.3}$   \\ \hline
 \end{tabular}
\caption{SM predictions and experimental values of $B\to K^*\mu^+\mu^-$ observables \cite{Mahmoudi:2012un}.
\label{tab:experiment}}
\end{center}
\end{table}


The decay $B\to K^*\mu^+\mu^-$ offers many other theoretically clean observables such as transverse amplitudes \cite{Egede:2010zc}, which will be measured in the near future.

\section{MFV results and predictions}
In the SM, the symmetry of the gauge interactions under the flavour group is only broken by the Yukawa couplings. Any NP model where all flavour and CP violating interactions can be linked to the known Yukawa couplings is said to be minimal flavour violating \cite{D'Ambrosio:2002ex}. The hypothesis of MFV provides a model independent solution to the NP flavour.

Compared to a model independent analysis (see for example Refs.\cite{Altmannshofer:2012ir,Bobeth:2011nj,DescotesGenon:2011yn}), the MFV hypothesis singles out a specific number of operators. Moreover, predictions based on the MFV benchmark have a specific meaning: Any measurement beyond the MFV bounds unambiguously indicates the existence of new flavour structures \cite{D'Ambrosio:2002ex}.

\begin{figure}[t!]
\begin{center}
\includegraphics[width=5.1cm]{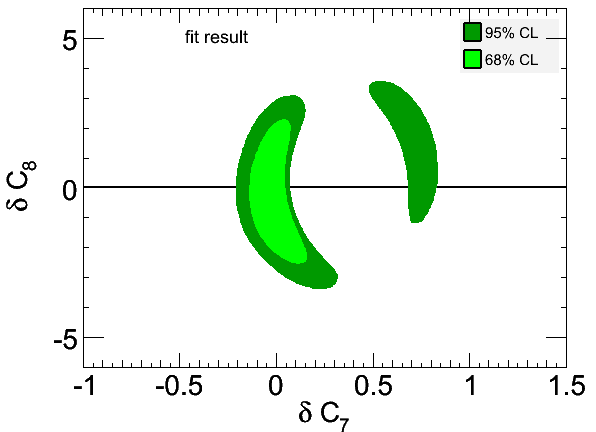}%
\includegraphics[width=5.1cm]{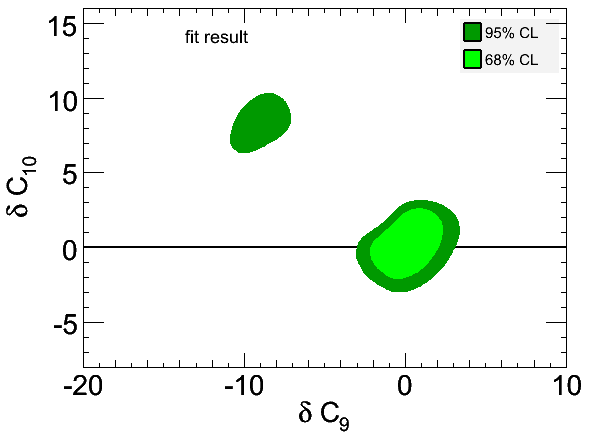}%
\includegraphics[width=5.1cm]{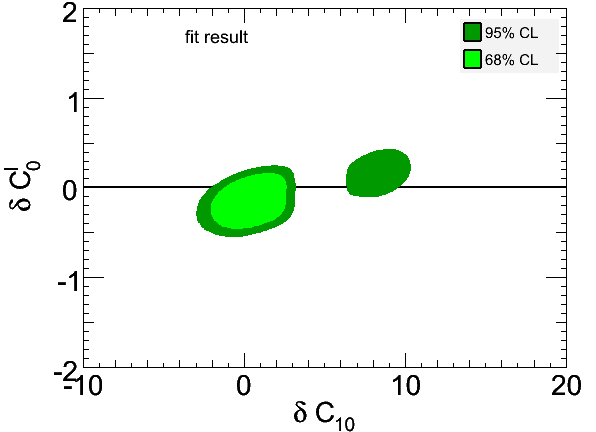}\\[0.3cm]
\includegraphics[width=5.1cm]{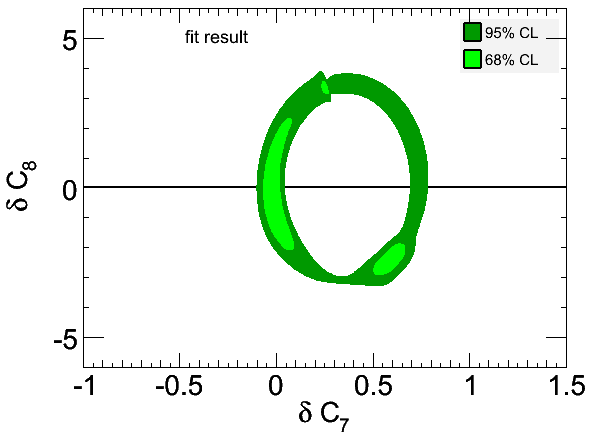}%
\includegraphics[width=5.1cm]{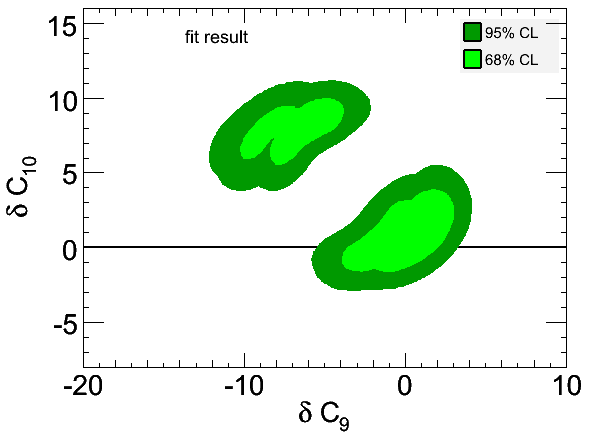}%
\includegraphics[width=5.1cm]{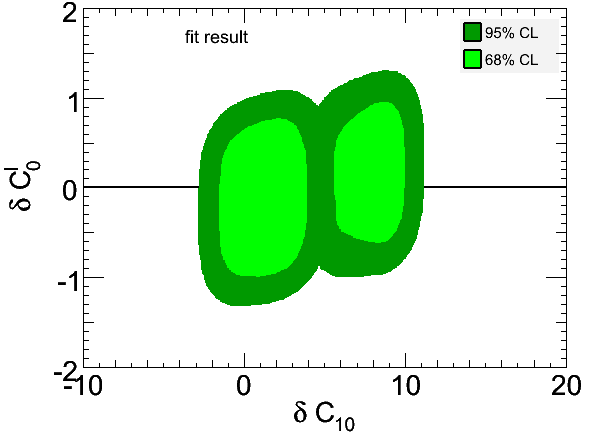}%
\caption{Global MFV fit to the various NP coefficients $\delta C_i$ in the MFV effective theory {\it with} (upper panel) and  {\it without} experimental data of LHCb (lower panel).}
\label{MFVfit}
\end{center}
\end{figure}

To derive the consequences of the MFV hypothesis we consider a set of $\Delta F = 1$ observables which includes in addition to the ones discussed in the previous section, the radiative decay $\bar B \to X_s \gamma$, isospin asymmetry $\Delta_0(B \to K^*\gamma)$ and inclusive $\bar B \to X_s \mu^+\mu^-$ decay as described in \cite{Hurth:2012jn}.
To obtain constraints on the Wilson coefficients, we scan over $\delta C_7$, $\delta C_8$, $\delta C_9$, $\delta C_{10}$ and $\delta C_0^l$ (where $C^l_{0}=2 C_{Q_1} = -2 C_{Q_2}$ in MFV). For each point we compute the flavour observables using {\tt SuperIso} and compare the predictions with the experimental results by calculating the $\chi^2$. The SM predictions and experimental results can be found in Tables~3 and 4 of \cite{Hurth:2012jn}. The global fit results are shown in Fig.~\ref{MFVfit}. To see the impact of the recent LHCb results, the consequences of the MFV fit prior to the LHCb data are also provided.

\begin{figure}[t!]
\begin{center}
\includegraphics[width=5.1cm]{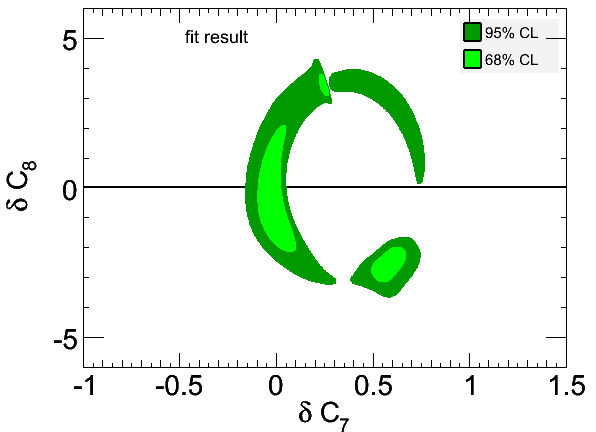}%
\includegraphics[width=5.1cm]{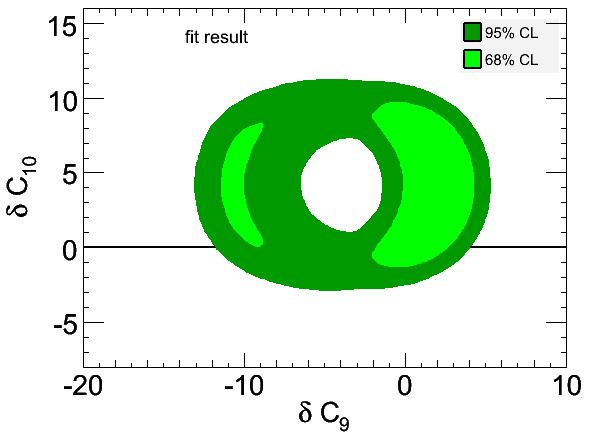}%
\includegraphics[width=5.1cm]{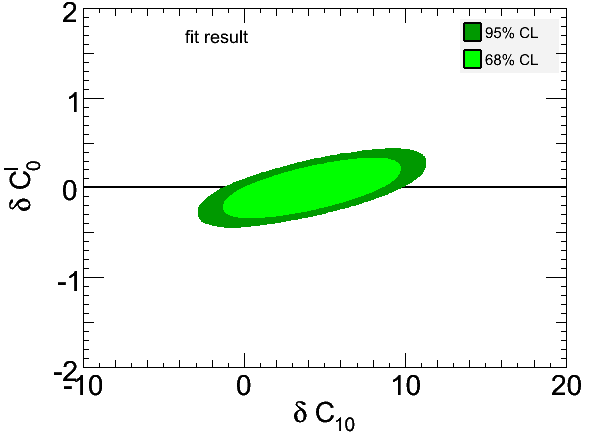}%
\caption{Global MFV fit with the latest data set {\it excluding}  all LHCb measurements of $B\to K^* \mu^+\mu^-$ observables.}
\label{MFVfitextra}
\end{center}
\end{figure}
In Fig.~\ref{MFVfitextra} we show the results when removing the LHCb measurements of $B\to K^* \mu^+\mu^-$ observables from the fit to pin down explicitly the effect of those measurements.

Using the allowed ranges for the Wilson coefficients obtained from the global fit we are in the position to make predictions for the observables which are not yet measured. The most important predictions are listed below:
\begin{itemize}
 \item $\mbox{BR}(B_d \to \mu^+ \mu^-) < 0.38 \times 10^{-9} $ (the current LHCb limit is $\mbox{BR}(B_d \to \mu^+ \mu^-) < 1.0 \times 10^{-9} $)\\[-8mm]

 \item $10^{-7} < {\rm  BR}(\bar B \to X_s \tau^+\tau^-)_{q^2>14.4 \rm{GeV}^2} <  3.7 \times 10^{-7}$\\[-8mm]

 \item $q^2_0(A_{FB}(B \to X_s \mu^+\mu^-)) > 1.94\, {\rm GeV}^2$\\[-8mm]

 \item $B \to K^* \mu^+ \mu^-$ transverse asymmetries:\\[-8mm]
\begin{itemize}
 \item $A_T^{(2)} \in [-0.065,-0.022]$ \hspace*{3.cm} $A_T^{(3)} \in [0.34,0.99]$\\[-6mm]
\item $A_T^{(4)} \in [0.19,1.27]$ \hspace*{3.97cm} $A_T^{(5)} \in [0.15,0.49]$
\end{itemize}

\end{itemize}

These predictions are very important as a measurement beyond these predictions would be a clear indication of a new flavour structure.

\section{SUSY constraints}
We now turn to the implications in SUSY, focusing on the CMSSM scenario, which assumes SUSY breaking mediated by gravity and is characterised by the set of parameters $[m_0,m_{1/2},A_0,\tan\beta$, $\mathrm{sign}(\mu)]$ defined by unification boundary conditions at the GUT scale.

\begin{figure}[t!]
\begin{center}
\includegraphics[width=5.5cm]{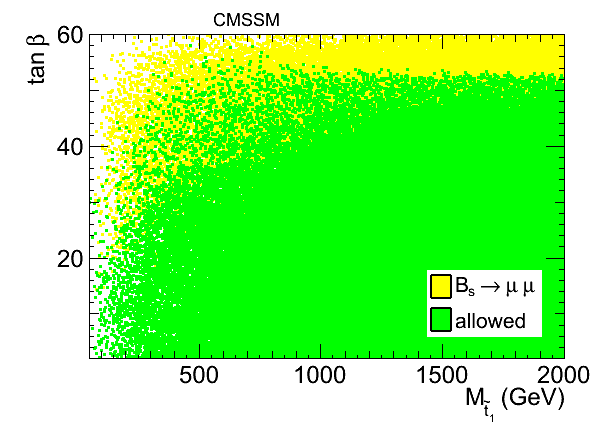}%
\includegraphics[width=5.5cm]{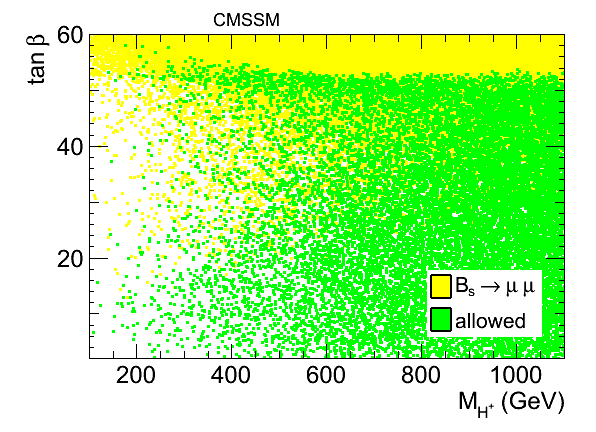}%
\caption{Constraints from $B_s \to \mu^+ \mu^-$ on the CMSSM, in the $(m_{\tilde{t}_1},\tan\beta)$ (left panel) and $(M_{H^+},\tan\beta)$ (right panel) parameter planes. The points compatible with the $B_s \to \mu^+ \mu^-$ are displayed on top.
\label{fig:bsmumu_cmssm}}
\end{center}
\end{figure}

The $B_s \to \mu^+ \mu^-$ decay receives large SUSY corrections which can enhance its branching ratio by orders of magnitude in the large $\tan\beta$ regime. Using the constraint at 95\% C.L. which includes theoretical uncertainties:
\begin{equation}
 {\rm BR}(B_s \to \mu^+ \mu^-) < 5.0 \times 10^{-9}\;,
\end{equation}
we investigate the consequences in the CMSSM by varying all the parameters in random scans. In Fig.~\ref{fig:bsmumu_cmssm}, we present the effect of this limit in the $(M_{\tilde{t}_1},\tan\beta)$ and $(M_{H^+},\tan\beta)$ parameter planes. We notice that $B_s \to \mu^+ \mu^-$ excludes the region with $\tan\beta \gtrsim 50$ independently of the stop and charged Higgs masses, while lower $\tan\beta$ values are only affected for light stop and charged Higgs masses~\cite{Akeroyd:2011kd}. 

\begin{figure}[t!]
\begin{center}
\includegraphics[width=5.cm]{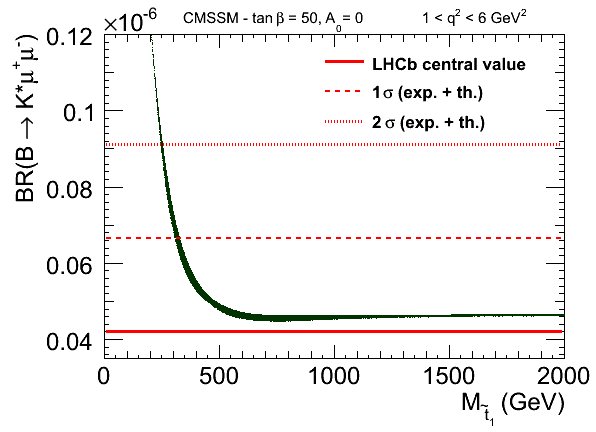}%
\includegraphics[width=5.cm]{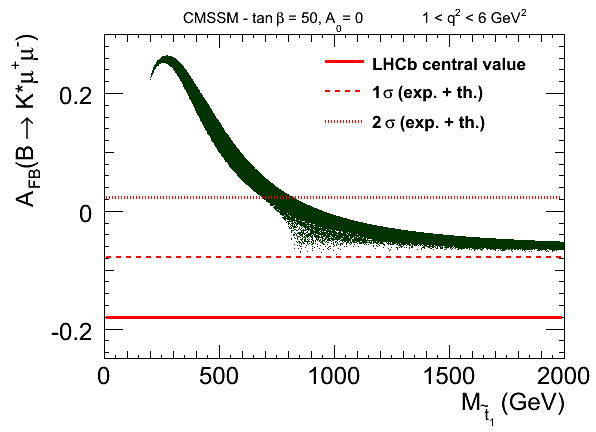}%
\includegraphics[width=5.cm]{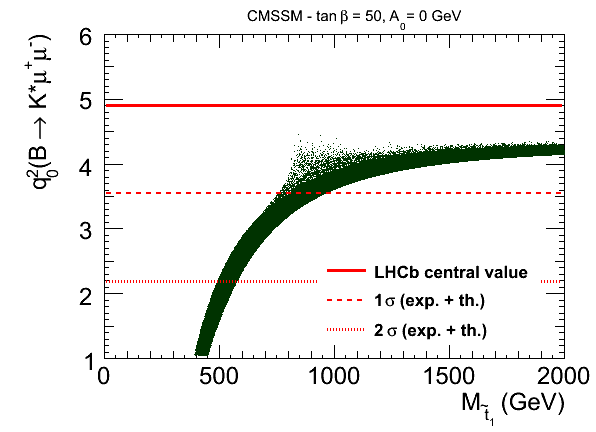}%
\caption{SUSY spread of the averaged ${\rm BR}(B\to K^*\mu^+\mu^-)$, $A_{FB}$ and zero crossing of $A_{FB}$, in the low $q^2$ region in function of the lightest stop mass, for $\tan\beta=50$ and $A_0=0$\label{fig:bkmumu_susy}.}
\end{center}
\end{figure}

The constraints from $B\to K^*\mu^+\mu^-$ observables have been thoroughly studied in \cite{Mahmoudi:2012un}. We consider here the case of $\tan\beta$=50, and investigate the SUSY spread as a function of the lightest stop mass. In Fig.~\ref{fig:bkmumu_susy}, we show the CMSSM results for $\tan\beta$=50 and $A_0=0$ for the averaged differential branching ratio at low-$q^2$, the forward-backward asymmetry $A_{FB}$ and the zero-crossing $q^2_0$ of $A_{FB}$. The solid red lines represent the LHCb central value, and the dashed and dotted lines the 1 and 2$\sigma$ bounds respectively. Both theoretical and experimental errors are considered and have been added in quadrature. We notice that the branching ratio excludes $M_{\tilde t_1} \lesssim$ 250 GeV for $\tan\beta$=50. The angular distributions in which the theoretical uncertainties are reduced could provide more robust constraints on the SUSY parameter space. As can be seen from the figure, $A_{FB}$ provides the most stringent constraints and excludes $M_{\tilde t_1} \lesssim$ 800 GeV at $\tan\beta$=50, while the zero crossing $q^2_0$ excludes $M_{\tilde t_1} \lesssim$ 550 GeV. The same observables in the high-$q^2$ region have less impact on the SUSY parameters.

\section{SuperIso program}
The {\tt SuperIso} program \cite{superiso} is a public C code dedicated to the computation of flavour physics observables.
Various models are implemented, and in particular SM, THDM, MSSM and NMSSM.
{\tt SuperIso} provides a broad set of flavour physics observables, which includes the branching ratio of $B \to X_s \gamma$, isospin asymmetry of $B \to K^* \gamma$, branching ratios of $B_{s,d} \to \mu^+ \mu^-$, branching ratios of $B \to X_s \ell^+ \ell^-$, $B \to K^* \mu^+ \mu^-$, as well as several angular quantities in these decays, branching ratio of $B_u \to \tau \nu_\tau$, branching ratio of $B \to D \tau \nu_\tau$, branching ratio of $K \to \mu \nu_\mu$, branching ratio of $D \to \mu \nu_\mu$, and the branching ratios of $D_s \to \tau \nu_\tau$ and $D_s \to \mu \nu_\mu$.
{\tt SuperIso} also provides the calculation of the anomalous magnetic moment of the muon. It uses a SUSY Les Houches Accord (SLHA) file \cite{slha} as input, which can be generated automatically by the program via a call to a spectrum generator or provided by the user. An extension of SuperIso which includes the calculation of relic density, {\tt SuperIso Relic}, is also publicly available \cite{superiso_relic}.
Finally, {\tt SuperIso} can provide output in the Flavour Les Houches Accord (FLHA) format \cite{Mahmoudi:2010iz}.

\section{Conclusion}

Over the past decades, flavour physics has established its role as an important player in the indirect search for new physics. This has now entered a new era with the start of the LHC and we presented here specific examples to demonstrate the implications of the recent LHCb results. 
We showed that the MFV hypothesis can be tested in a generic way. The MFV predictions can be checked in the near future and any measurement beyond the predictions would clearly indicate new flavour structure.

The high priority LHCb processes, namely $B_s \to \mu^+ \mu^-$ and $B \to K^* \mu^+ \mu^-$ have impressive impacts both for model independent analysis and specific NP scenarios such as supersymmetry as discussed here. The theoretical uncertainties are under control for both decays and
with more data becoming available these decays will play an even more important role in constraining NP scenarios in the near future.

\end{document}